\documentclass[11pt]{article}
\usepackage[dvipsnames]{xcolor}

\usepackage[framemethod=default]{mdframed}
\usepackage{caption}
\usepackage{indentfirst}
\usepackage{bm, mathrsfs, graphics,float,amssymb,amsmath,subeqnarray,setspace,graphicx,amsthm,epstopdf,subfigure, enumerate, color}
\usepackage[utf8]{inputenc}
\usepackage[colorlinks,
            linkcolor=red,
            anchorcolor=blue,
            citecolor=blue
            ]{hyperref}
\usepackage{natbib}
\usepackage{fullpage}

\usepackage{multirow}
\usepackage{tablefootnote}
\usepackage{colortbl}% http://ctan.org/pkg/xcolor
\usepackage{hhline}% http://ctan.org/pkg/hhline
\usepackage[titletoc]{appendix}
\usepackage{algorithm}
\usepackage{algorithmic}
\usepackage{mathtools}

\begin{document}
\title{
Compressing LSTM Networks by Matrix Product Operators
}

\author{
Ze-Feng Gao
\thanks{
Renmin University of China;
email: \{zfgao, lgao, zlu\}@ruc.edu.cn
}
\qquad
Xingwei Sun
\thanks{
Institute of Acoustics, Chinese Academy of Sciences and University of Chinese Academy of Sciences;
email:\{sunxingwei,lijunfeng\}@hccl.ioa.ac.cn
}
\qquad
Lan Gao
\footnotemark[1]
\qquad
Junfeng Li
\footnotemark[2]
\qquad
Zhong-Yi Lu
\footnotemark[1]
\thanks{
	Corresponding author
}
}
\date{}

\maketitle
%\tableofcontents

\begin{abstract}
Long Short Term Memory~(LSTM) models are the building blocks of many state-of-the-art natural language processing~(NLP) and speech enhancement~(SE) algorithms. 
% Neural networks built on LSTM have achieved remarkable results in many fields. 
However, there are a large number of parameters in an LSTM model. This usually consumes a large number of resources to train the LSTM model. Also, LSTM models suffer from computational inefficiency in the inference phase. 
Existing model compression methods (e.g., model pruning) can only discriminate based on the magnitude of model parameters, ignoring the issue of importance distribution based on the model information.
Here we introduce the MPO decomposition, which describes the local correlation of quantum states in quantum many-body physics and is used to represent the large model parameter matrix in a neural network, which can compress the neural network by truncating the unimportant information in the weight matrix.
In this paper, we propose a matrix product operator~(MPO) based neural network architecture to replace the LSTM model. The effective representation of neural networks by MPO can effectively reduce the computational consumption of training LSTM models on the one hand, and speed up the computation in the inference phase of the model on the other hand. We compare the MPO-LSTM model-based compression model with the traditional LSTM model with pruning methods on sequence classification, sequence prediction, and speech enhancement tasks in our experiments. The experimental results show that our proposed neural network architecture based on the MPO approach significantly outperforms the pruning approach.
\end{abstract}
\section{Introduction}
The Long Short-Term Memory~(LSTM) model~\cite{hochreiter1997long} has become a popular choice for modeling many practical tasks, such as speech recognition~\cite{xiong2016achieving}, language modeling~\cite{jozefowicz2016exploring, shazeer2017outrageously, sundermeyer2012lstm}, machine translation~\cite{wu2016google}, and many other tasks. These temporal and sequential modelings show that many state-of-the-art results~\cite{he2017deep, lee2017end, seo2016bidirectional,peters2018deep} have been achieved by the LSTM model.

However, the scalability of LSTM models has a significant drawback in that most LSTM models have a large number of parameters, which can lead to high computational costs for training and using neural networks based on LSTM modules.
Since LSTM models usually consist of multiple linear and nonlinear transformations, multiple high-dimensional matrices are required to represent these parameters. In one time step, we need multiple linear transformations between the dense matrix of high-dimensional inputs and the hidden states of the previous time step. Especially in the field of speech recognition and machine translation, the latest models require a large computational cost with millions of parameters, which can only be implemented in high-end cluster environments~\cite{schuster2010speech}. 
This prevents efficient LSTM models from being fast enough for large-scale real-time inference or small enough to be implemented in low-end devices with limited memory~(\emph{e.g.}, cell phones or embedded systems).
Moreover, although the storage capacity of an LSTM model is considered to be proportional to the size of the model, recent research has proved the opposite fact, suggesting that an LSTM model is indeed over-parameterized~\cite{levy2018long,melis2017state,merity2017regularizing,chen2022energy,ma2022knowledge}.
Most of the existing works are based on parameter pruning~\cite{jiang2022low}, model distillation~\cite{gupta2022compression} and model quantization methods~\cite{chen2022energy} to compress LSTM models. However, these methods rely on strong prior knowledge or large computational resources~\cite{gupta2022compression}, and also impair model performance.

In this work, we propose an MPO-LSTM model, which is an MPO network~\cite{gao2020compressing} based on the LSTM model architecture. 
An important merit of MPO decomposition is that the parameters can be reordered according to the different importance of information. A series of local tensors can be obtained after MPO decomposition, and the effective compression of the linear part can be achieved by truncating the connection keys between the local tensor.
Specifically, we apply the MPO format to reformulate two dense matrices in the LSTM model structure, one is the dense matrix between the input vector and hidden layer, the other is the dense matrix between the dense matrices of the high dimensional input and the previous hidden state in a time-step.

In the experiment section, we evaluate the proposed MPO-LSTM model on sequence classification, sequence prediction, and speech enhancement tasks. 
Next, we compare the results of the MPO-LSTM model with those of the compressed LSTM model using the pruning method with the same number of parameters. The experimental results show that our method has more advantages than pruning.

In summary, our contributions are as follows:

\begin{itemize}
	\item We demonstrate that deep neural networks can be well compressed using matrix product operators with proper structure and balanced dimensions, at the same time without performance degradation.
% 	\item We propose a novel compression method based on the MPO that can drastically compress an LSTM model while learning. 
	\item We propose a new network structure using the MPO method, and show that we can experimentally achieve even higher accuracy than the pruning method at the same compression ratio in natural language processing and acoustic fields.
	
\end{itemize}

\section{Priliminary}\label{mpo-pruning-lstm}

\subsection{Long Short-Term Memory}

To learn long-range dependencies with Recurrent Neural Network (RNN) is challenging due to the vanishing and exploding gradient problems \cite{bengio1994learning,pascanu2013difficulty}.
To address this issue, the LSTM model has been introduced by~\cite{hochreiter1997long}, with the following recurrent computations:
\begin{equation}
    LSTM: h_{t-1},c_{t-1},x_t\rightarrow h_t,c_t. \label{Eq:LSTM-1}
\end{equation}

Here $x_t$ is an input vector, $h_t$ is the cell state, $c_t$ is the cell memory. Equation ($\ref{Eq:LSTM-1}$) is computed as follows,

\begin{equation}
\begin{split}
{\begin{pmatrix}
i_t      \\f_t      \\o_t   \\
{\hat{c}_t}
\end{pmatrix}} &=
{\begin{pmatrix}
\sigma      \\\sigma      \\\sigma   \\tanh
\end{pmatrix}}
(W_i \ W_h)
{\begin{pmatrix}
x_t     \\h_{t-1}
\end{pmatrix}} \\
\end{split}
\label{eq:LSTM-eq2}
\end{equation}
\begin{equation}
W_i =
{\begin{pmatrix}
W_i^i     \\W_i^f     \\W_i^o     \\W_i^c
\end{pmatrix}
},
W_h =
{\begin{pmatrix}
W_h^i \\W_h^f  \\W_h^o     \\W_h^c
\end{pmatrix}
},
\label{eq:LSTM-eq3}
\end{equation}
\begin{eqnarray}
    c_t &=& f_t \odot c_{t-1} + i_c \odot \hat{c}_t, \nonumber \\
    h_t &=& o_t \odot tanh(c_t),
\end{eqnarray}
where $x_t \in R^{N_x}$ and $h_t \in R^{N_h}$ at time $t$. In the above equation, $\sigma(\cdot)$ and $\odot$ denote the sigmoid function and element-wise multiplication operator respectively. The $i_t, f_t, o_t, and ~\hat{c}_t$ are respectively the input gates, the forget gates, the output gates, and the memory cells.
The input gates retain the candidate memory cell values that are useful for the current memory cell and the forget gates retain the previous memory cell values that are also useful for the current memory cell. The output gates retain the memory cell values that are useful for the output and the next time-step hidden layer computation.

The major part of computational cost is in $W_i$ and $W_h$, where $W_i \in R^{N_x\times 4N_h}$ that is the input matrix, and $W_h \in R^{N_h \times 4N_h}$ that is the hidden matrix in time-step.

\subsection{Pruning method}
The pruning is the most commonly used sparsity method in the original domain. \cite{han2015learning,han2015deep} recursively trained a neural network and pruned unimportant connections based on their weight magnitudes. \cite{guo2016dynamic} proposed the dynamic network surgery prune and spliced the branch of the network.

In the pruning method, for the weight matrix between layers, some unimportant weights are cut off, usually by setting their values to zero.
In training, one can implement pruning by filtering the values of weights.
After pruning, the weight matrix of neural network has only the important weight parts left, and the corresponding, unimportant weight parts are trimmed away.

\section{Method}

\subsection{MPO Decomposition on Weight Tensor}\label{MPO weight}

The MPO method develops from quantum many-body physics, which is based on high order tensor single value decomposition.
An MPO is used to factorize a higher-order tensor into a sequential product of the so-called local-tensors, that is a more generalized form of the tensor-train approximation~\cite{verstraete2004matrix}.
By representing the linear transformations in a model with MPOs, the number of parameters needed is greatly shrunk since the number of parameters contained in an MPO decomposition format just grows linearly with the system size \cite{poulin2011quantum}.
The MPO method has been demonstrated to be well effective for model compression by using the MPO to replace the linear transformations of fully-connected and convolutional layers \cite{gao2020compressing}.

To clarify the MPO method process, we assume a weight matrix $W_{yx} \in \mathbf{R}^{N_x\times N_y}$, and then reshape it to a 2n-indexed tensor
\begin{equation}
  W_{yx} = W_{j_1j_2\cdots j_n,i_1i_2\cdots i_n} .  \label{Eq:IndexDecompose}
\end{equation}
As Eq.(\ref{Eq:IndexDecompose}) shows, each entry vector is reshaped as a sequence of matrix multiplications in which dimension $N_{x}$ is reshaped into a coordinate in an n-dimensional space, labelled by ${(i_{1}i_{2} \cdots i_{n})}$. Hence, there is one-to-one mapping between input vector $X$ and MPO label ${(i_{1}i_{2} \cdots i_{n})}$. Likewise, we can set up another one-to-one correspondence between $Y$ and ${(j_{1}j_{2} \cdots j_{n})}$. If ${I_{k}}$ and  ${J_{k}}$ are the dimensions of index ${i_{k}}$ and ${j_{k}}$ respectively,
\begin{equation}
    \prod_{k=1}^{n}I_{k} = N_{x}, \quad \prod_{k=1}^{n}J_{k} = N_{y}  . \label{eq:dims}
\end{equation}
The MPO representation of $W$ is obtained by factorizing it into a product of $n$ local-tensors( Usually called core-tensors).

\begin{eqnarray}
  W_{j_1\cdots j_n,i_1\cdots i_n} = \mathrm{Tr} \left( w^{(1)} [j_1,i_1]w^{(2)} [j_2,i_2] \cdots w^{(n)} [j_n,i_n]  \right), \label{Eq:MPO}
\end{eqnarray}
where $w^{(k)}[j_k,i_k]$ is a $d_{k-1}\times d_{k}$ matrix, and the $d_k$ means the dimension on the bond linking $w^{(k)}$ and $w^{(k+1)}$ with $d_0=d_n=1$.\\
\begin{figure}[ht]
\centering
\includegraphics[width=9cm]{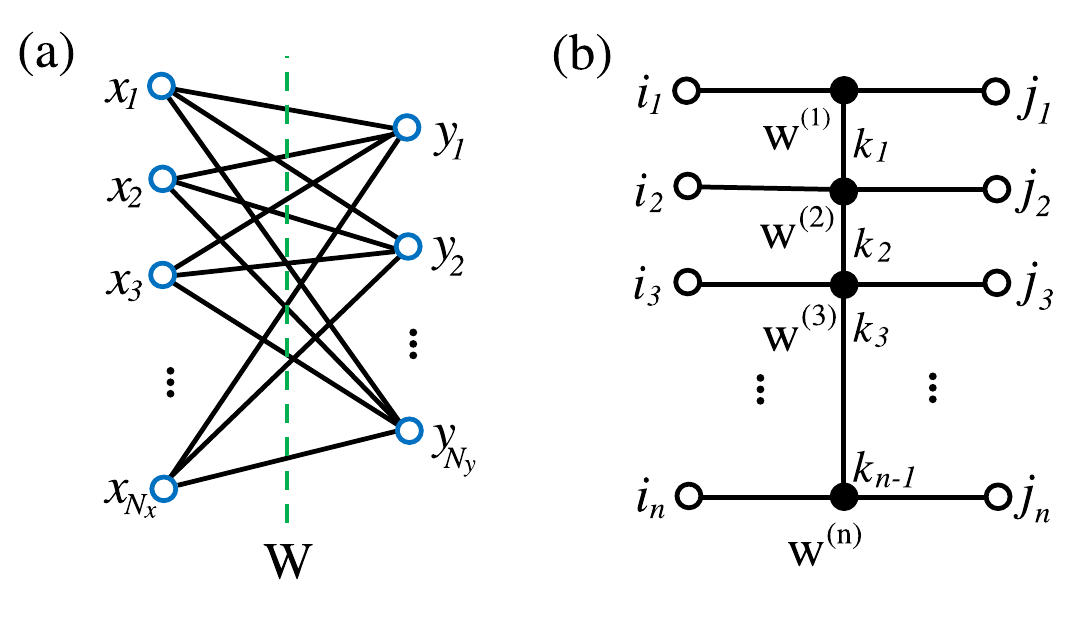}
  \caption{(a) Graphical representation of weight matrix $W$ in a fully connected layer. The blue circles represent neurons. The solid line connecting input neuron $x_i$ with output neuron $y_j$ represents weight element $W_{ji}$.
  (b) The weight matrix represented by MPO. The local operator tensors $w^{(k)}$ are represented by filled circles. The hollow circles denote the input and output indices, $i_l$ and $j_l$, respectively. Given $i_k$ and $j_k$, $w^{(k)}[j_k,i_k]$ is a matrix.
  }
\label{Fig:MPO}
\end{figure}

Any matrix can be represented with the MPO representation. We schematically show this representation  in Fig.\ref{Fig:MPO}. Such a matrix product structure leads to the fact that the scaling of the parameter number is reduced from exponential to polynomial, which is a great advantage of MPO representation. To be specific, the parameter number is shrunk with the following equation.
\begin{equation}
   \prod_{k=1}^{n}I_k J_k \rightarrow \sum_{k=2}^{n-1}I_k J_k d_{k-1}d_k + I_1J_1d_0 + I_nJ_nd_{n-1}   . \label{eq:reduced_D}
\end{equation}
In Eq.(\ref{eq:reduced_D}), $d_k$ is dubbed as bond dimension. In practice, $d_k$ can be regarded as a tunable parameter which controls the accuracy of the representation, i.e., the more larger $d_k$ is, the more parameters the MPO contains.
We discuss the issue in detail in Section 4.2 and explain it in Figure \ref{Fig:Parameter}.

\begin{table}[h]%[htbp]
\centering
\renewcommand \arraystretch{1.1}
\begin{tabular}{|c|c|c|}   \hline
     {Operation}     &  {Time}                           & {Memory}
     \\ \hline
     {FC forward}    & {$O(N_x N_y$)}                    & {$O(N_x N_y$)}
     \\ \hline
     {MPO forward}   & {$O(nd^2 m$ $ max(N_x, N_y)$)}     & {$O(d$ $max(N_x, N_y)$)}
     \\ \hline
     {FC backward}   & {$O(N_x N_y$)}                    & {$O(N_x N_y$)}
     \\ \hline
     {MPO backward}   & {$O(n^2d^4 m$ $ max(N_x, N_y)$)}     & {$O(d^3$ $max(N_x, N_y)$)}
     \\ \hline
\end{tabular}
\caption{Full-Connected Layer And MPO Layer Running Time And Memory. In this table, $n$ denotes the number of MPO core tensors, $m$ denotes $max(\{I_k\}_{k=1}^{n})$, $d$ denotes $max(\{d_k\}_{k=0}^{n})$}, $N_x$ denotes the total dimension of input, $N_y$ denotes the total dimension of output, respectively.
\label{table:MPO-time-memory}
\end{table}
Additionally, Tabel \ref{table:MPO-time-memory} compares the forward and backward propagation times and the memory complexity between the fully connected layer and the MPO layer in Big-O notation proposed by \cite{novikov2015tensorizing}. We compare the fully connected layer with matrix $W \in R^{N_x\times N_y}$  versus the MPO layer in format $MPO(W,x)$ with MPO-dimension $\{d_k\}_{k=0}^{n}$. 
As can be seen from the table, the MPO format has more advantages than the traditional one in terms of time and memory consumption.

\subsection{MPO Long-Short Term Memory and motivation}\label{mpo-c-lstm}
The tendency of LSTM model overfitting suggests that there is always redundancy among the weights. Inspired by the low-rank decomposition of weight matrices in \cite{gao2020compressing} and \cite{denil2013predicting} which can well reduce the model size and computational cost at the same time, we adopt a more generalized operators (MPO) method to replace all the linear parts of the LSTM model.

In this work, we investigate the speech enhancement of several typical cases with the LSTM model. Accordingly, we factorize input-to-hidden weight matrix $W$ with $MPO(W,x)$ and represent hidden-to-hidden weight matrix $U$ in the $MPO(U,h)$.\\
MPO-LSTM:
\begin{eqnarray}
    &k^{[t]}&= \sigma(MPO(W^{k},x^{[t]})+MPO(U^{k},h^{[t-1]})+b^{k}), \nonumber \\
    &f^{[t]}&= \sigma(MPO(W^{f},x^{[t]})+MPO(U^{f},h^{[t-1]})+b^{f}), \nonumber \\
    &o^{[t]}&= \sigma(MPO(W^{o},x^{[t]})+MPO(U^{o},h^{[t-1]})+b^{o}), \nonumber \\
    &g^{[t]}&= tanh(MPO(W^{g},x^{[t]})+MPO(U^{g},h^{[t-1]})+b^{g}), \nonumber \\
    &c^{[t]}&= f^{[t]} \circ c^{[t-1]} + k^{[t]} \circ g^{[t]}, \nonumber \\
    &h^{[t]}&= o^{[t]} \circ tanh(c^{[t]}),
\end{eqnarray}
We can see that to construct an MPO-LSTM model we need to prepare eight MPOs, one for each of the gating units and hidden units respectively. Instead of calculating these MPOs directly, we increase the dimension of the first tensor to form the output tensor. This trick, inspired by the implementation of standard LSTM model in \cite{chollet2015keras}, can further reduce the number of parameters, where the concatenation is actually participating in the tensorization. The compression ratio for input-to-hidden weight matrix $W$ now becomes
\begin{equation}
    \rho_{w} = \frac{\sum_{k=1}^{d}i_{k}j_{k}d_{k-1}d_{k} + 3\cdot(i_{1}n_{1}d_{0}d_{1})}{4\cdot\prod_{k=1}^{d}i_{k}n_{k}}
    \label{eq:parameter}
\end{equation}
Meanwhile, the compression ratio hidden-to-hidden weight matrix $U$ becomes:
\begin{equation}
    \rho_{u} = \frac{\sum_{k=1}^{d}i_{k}^{'}j_{k}^{'}d_{k-1}^{'}d_{k}^{'} + 3\cdot(i_{1}^{'}j_{1}^{'}d_{0}^{'}d_{1}^{'})}{4\cdot\prod_{k=1}^{d}i_{k}^{'}j_{k}^{'}}
\end{equation}
Thus, the total compression ratio is
\begin{eqnarray}
    \rho_{*}=&&\frac{\sum_{k=1}^{d}i_{k}j_{k}d_{k-1}d_{k} + 3\cdot(i_{1}n_{1}d_{0}d_{1})+
    \sum_{k=1}^{d}i_{k}^{'}j_{k}^{'}d_{k-1}^{'}d_{k}^{'} + \cdot(i_{1}^{'}j_{1}^{'}d_{0}^{'}d_{1}^{'})}{4\cdot\prod_{k=1}^{d}i_{k}n_{k}+4\cdot\prod_{k=1}^{d}i_{k}^{'}j_{k}^{'}} 
\end{eqnarray}
In a specific MPO, the structure is variational. In our calculation, for simplicity all $d_{k}$ are set equal in the same MPO method structure, and denoted as $d$, as we met before. The index decomposition is not unique, and in this work, we just choose by convenience.

\section{Experiments}\label{experiments}

In this section, we evaluate our proposed LSTM model with MPO method (MPO-LSTM) and compare it with baseline LSTM model and pruning-LSTM model under the same compression rate. Specifically we compare our method with the pruning method for model compression rates from 5 to 100. In our experiments, the models with pruning method are referred as pruning-LSTM and the models with MPO method are refered as MPO-LSTM.

To evaluate and compare the performance of the MPO method with the pruning method, we conducted the experiments on the sentiment analysis classification tasks, in which we predicted whether the sequence of tokens(usually words or sentences) contains either positive or negative meaning, and the sequence regression tasks, in which we predicted the clean speech from the noisy speech. 
We used Internet Movie Database(IMDB) and Stanford Sentiment Treebank(SST) datasets for the sentiment analysis classification task, and we used the speech enhancement datasets VoiceBank-DEMAND (VBD) for the sequence regression tasks \cite{valentini2016investigating}.
For all of tasks, we adopted Adam \cite{kingma2014adam} to optimize our model parameters.

\subsection{Sentiment analysis classification on IMDB and SST}
We evaluated our proposed  MPO-LSTM model and the pruning-LSTM model for the classification task using the IMDB dataset \cite{maas2011learning} and the Stanford Sentiment Treebank (SST) \cite{socher2013recursive} with five categories. The IMDB dataset consists of 50,000 biased comments for two classes that are either positive or negative. The IMDB dataset has a training set with 25,000 images, and a test set with 25,000 images. 
We took the most frequent 25,000 words for the IMDB dataset and 17,200 for SST respectively, then embedded them into a standard embedding layer and performed classification respectively using the pruning-LSTM model and the MPO-LSTM model both with hidden size $h$. In our experiments, we set $h$ to 256.

As shown in Section 3.2, there are two weight matrices, namely input-to-hidden weight matrix $W$ and hidden-to-hidden weight matrix $U$. In the MPO-based compression method, both weight matrices are decomposed. Accordingly, the parameters including factorization factors $i_k^W, j_k^W, i_k^U, j_k^U$ and bond dimension factors $d_k^W, d_k^U$ are adjustable. In these tasks, we fixed the factorization factors as $(8,2,2,8) \times (8,2,2,8)$ and adjusted the bond dimension factors for the two weight matrices to achieve a given compression rate. The values of the tunable bond dimension at different compression rates are listed in Table \ref{tab:dimension-lstm}. In the pruning method, we only need to determine the sparsity of the weight matrix.
\begin{table}
\centering
\renewcommand \arraystretch{1.1}
\begin{tabular}{|l|l|l|l|l|l|l|l|l|}
\hline
$\rho_{*}$    & 5  & 10 & 15 & 20 & 25 & 50 & 75 & 100 \\ \hline
$d^W$ & 64 & 41 & 32 & 26 & 22 & 13 & 9  & 7   \\ \hline
$d^U$ & 64 & 40 & 29 & 24 & 20 & 13 & 9  & 7   \\ \hline
\end{tabular}
\caption{In IMDB and SST-5 tasks, the factorization factors  adopted as $(8,2,2,8) \times (8,2,2,8)$ and the bond dimension factors adjusted for the two weight matrices to achieve a given compression rate. The $\rho_{*}$ denotes the total compression ratio of neural network, the $d^W$ denotes bond dimension of the matrix mapping from the input-to-hidden weight matrix, the $d^U$ denotes bond dimension of the matrix mapping from the hidden-to-hidden weight matrix, respectively.}
\label{tab:dimension-lstm}
\end{table}

\begin{figure}[ht]
\centering
\includegraphics[width=15cm]{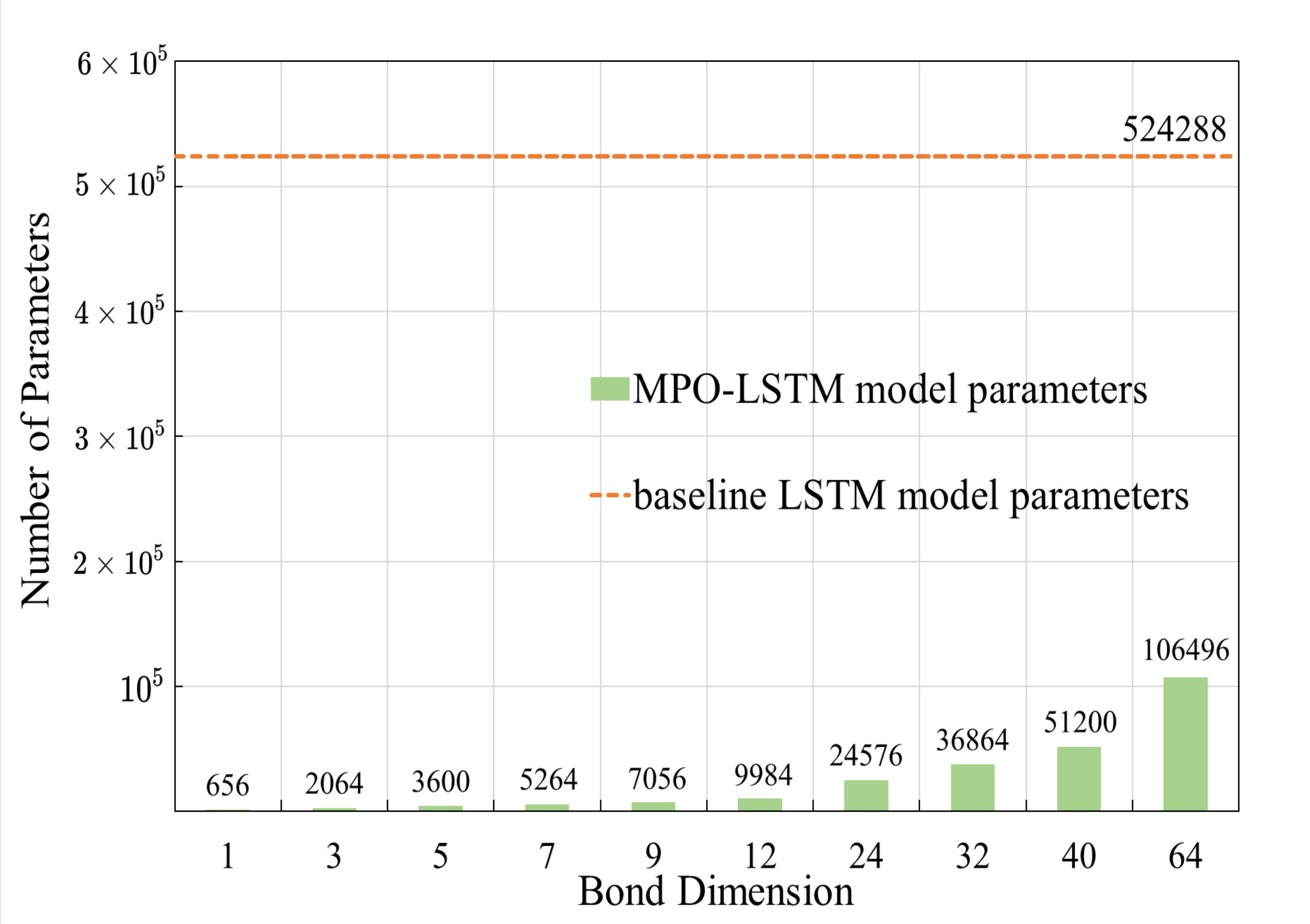}
  \caption{The number of parameters w.r.t bond dimension $d$ of MPO-LSTM model, in the setting of $I = 256, J = 256$. While the baseline  LSTM model contains 524288 parameters. The number parameter of MPO-LSTM model is in growth with $d$, meanwhile, the total number is always smaller than those of the baseline model. 
  }
\label{Fig:Parameter}
\end{figure}

We show the parameters of a single MPO-LSTM model as the bond dimension changes in Fig.\ref{Fig:Parameter}. We can see that with the bond dimension of the MPO-LSTM model becoming larger, the parameters also increase. When we set $d = 64$, although the model parameters increase, they are still much less than those in the baseline LSTM model.\\
Our findings are summarized in Table \ref{tab:lstm-imdb} and Table \ref{tab:lstm-sst}. We observe that as the compression rate becomes larger, the accuracy of MPO method does not decrease, and is higher than that of the baseline LSTM model, indicating that there are many redundant parameters in the original LSTM model. Thus we need effective methods, like MPO based method, to reduce these redundant parameters. 
\begin{table}[!htb]
    \centering
    \renewcommand{\arraystretch}{1.1}
        \begin{tabular}{|c|c|c|}\hline
            \multirow{2}{*}{\begin{tabular}[c]{@{}c@{}}Compression\\ Rate\end{tabular}} & \multirow{2}{*}{\begin{tabular}[c]{@{}c@{}}Compression\\ Method\end{tabular}} & \multirow{2}{*}{\begin{tabular}[c]{@{}c@{}}Test Accuracy(\%)\end{tabular}} \\
            &  & \\ \hline
            0 & - &  88.30 \\ \hline
            \multirow{2}{*}{5} & Pruning & 87.02  \\
             & MPO & \textbf{88.33}  \\ \hline
            \multirow{2}{*}{10} & Pruning &  86.57 \\
            & MPO &  \textbf{88.52} \\ \hline
            \multirow{2}{*}{15} & Pruning & 87.42 \\
            & MPO & \textbf{88.14}  \\ \hline
            \multirow{2}{*}{20} & Pruning & 86.59 \\
            & MPO  & \textbf{88.35}  \\ \hline
            \multirow{2}{*}{25} & Pruning & 86.76 \\
            & MPO & \textbf{88.51} \\ \hline
            \multirow{2}{*}{50} & Pruning & 87.68 \\
            & MPO & \textbf{88.44} \\ \hline
            \multirow{2}{*}{75} & Pruning & 87.21 \\
            & MPO & \textbf{88.57} \\ \hline
            \multirow{2}{*}{100} & Pruning & 87.11 \\
            & MPO &  \textbf{88.64} \\ \hline
        \end{tabular}
    \caption{The Test Accuracy results with IMDB Dataset of the LSTM model respectively with pruning and MPO base Methods.}
    \label{tab:lstm-imdb}
\end{table}
\begin{table}[!htb]
    \centering
    \renewcommand{\arraystretch}{1.1}
    \begin{tabular}{|c|c|c|}\hline
        \multirow{2}{*}{\begin{tabular}[c]{@{}c@{}}Compression\\ Rate\end{tabular}} & \multirow{2}{*}{\begin{tabular}[c]{@{}c@{}}Compression\\ Method\end{tabular}} & \multirow{2}{*}{\begin{tabular}[c]{@{}c@{}}Test Accuracy(\%)\end{tabular}} \\
        &  & \\ \hline
        0 & - &  44.10 \\ \hline
        \multirow{2}{*}{5} & Pruning & 41.16  \\
          & MPO & \textbf{43.09}  \\ \hline
        \multirow{2}{*}{10} & Pruning &  41.59 \\
          & MPO &  \textbf{41.60} \\ \hline
        \multirow{2}{*}{15} & Pruning & 41.67 \\
          & MPO & \textbf{42.14}  \\ \hline
        \multirow{2}{*}{20} & Pruning & 41.38 \\
          & MPO  & \textbf{41.59}  \\ \hline
        \multirow{2}{*}{25} & Pruning & 41.45 \\
          & MPO & \textbf{42.05} \\ \hline
        \multirow{2}{*}{50} & Pruning & 41.29 \\
          & MPO & \textbf{41.77} \\ \hline
        \multirow{2}{*}{75} & Pruning & 41.15 \\
          & MPO & \textbf{41.95} \\ \hline
        \multirow{2}{*}{100} & Pruning & 41.09 \\
          & MPO &  \textbf{43.05} \\ \hline
        \end{tabular}
    \caption{The Test Accuracy results with SST Dataset of the LSTM model respectively with pruning and MPO base Methods.}
    \label{tab:lstm-sst}
\end{table}

In addition, we observe that the models heavily compressed by the MPO method can perform equally or even better than the uncompressed models. At various compression rates, the results of the MPO method are better than those of the pruning method. This shows that the MPO method can be applied as a simple and efficient alternative pruning method.

\subsection{ Speech enhancement results}
In the speech enhancement task, we used the LSTM model to estimate the ideal ratio mask (IRM) from several acoustic features for denoising \cite{wang2013exploring}. In our experiments, the VBD dataset was used in which 30 speakers selected from Voice Bank corpus \cite{veaux2013voice} were mixed with 10 noise types: 8 from Demand dataset \cite{thiemann2013diverse} and 2 artificially generated one. 
The test set was generated with 5 noise types from demand that did not coincide with those for training data. Consequently, 11,572 and 824 noisy-clean speech pairs were provided as the training and test set, respectively. This dataset is openly available and frequently used in experiments of DNN-based speech enhancement.

In the MPO method based LSTM model compression, we used the same network structure as the one in the sentiment analysis classification task, except the unit numbers of the input and output layers. In this speech enhancement model, the input and output layers both had 256 hidden units, which were the same as the dimension of the input feature and training target. To evaluate the speech enhancement performances of the LSTM models respectively with and without compression and further compare the two different compression methods, 
we adopted an objective measure proposed by \cite{Rix2002Perceptual}, namely the  perceptual evaluation of speech quality (PESQ).

The speech enhancement evaluation results of the LSTM models are shown in Table \ref{tab:lstmirm}.
These results confirm the effectiveness of the LSTM model in speech enhancement task. We can see that the LSTM model without compression gains 0.52 PESQ improvements on average of all the utterances in the test set in comparision with the noisy speech. In terms of compressed models, the speech enhancement performance decreases with the increase of compression rate for both compression methods. However, the MPO-based compression method performs better than the pruning method at the same compression rates. Even in the high compression rate case such as 100, the compressed model with the MPO-based method can still achieve satisfactory performance with only 0.17 PESQ loss.

 \begin{table}[!htb]
        \centering
        \renewcommand{\arraystretch}{1.1}
            \begin{tabular}{|c|c|c|}\hline
                \multirow{2}{*}{\begin{tabular}[c]{@{}c@{}}Compression\\ Rate\end{tabular}} & \multirow{2}{*}{\begin{tabular}[c]{@{}c@{}}Compression\\ Method\end{tabular}} & \multirow{2}{*}{\begin{tabular}[c]{@{}c@{}}PESQ (MOS)\end{tabular}} \\
                &  & \\ \hline
                0 & - & 2.50 \\ \hline
                \multirow{2}{*}{5} & pruning &2.44  \\
               & mpo & \textbf{2.45}  \\ \hline
               \multirow{2}{*}{10} & pruning &  2.42 \\
               & mpo &  \textbf{2.46} \\ \hline
              \multirow{2}{*}{15} & pruning & 2.37 \\
               & mpo & \textbf{2.39}  \\ \hline
              \multirow{2}{*}{20} & pruning & 2.35 \\
               & mpo  & \textbf{2.38}  \\ \hline
              \multirow{2}{*}{25} & pruning & 2.31 \\
               & mpo & \textbf{2.41} \\ \hline
              \multirow{2}{*}{50} & pruning & 2.25 \\
               & mpo & \textbf{2.38} \\ \hline
              \multirow{2}{*}{75} & pruning & 2.28 \\
               & mpo & \textbf{2.35} \\ \hline
              \multirow{2}{*}{100} & pruning & 2.33 \\
               & mpo &  \textbf{2.33} \\ \hline
                \multicolumn{2}{|c|}{Noisy Speech} &1.98 \\ \hline
            \end{tabular}
        \caption{The speech enhancement performance evaluation results of the LSTM model with pruning and MPO-base Methods.}
        \label{tab:lstmirm}
\end{table}

\section{Related Work}\label{related-work}
Our method differs from the existing methods in two folds. First, we propose a model called MPO-LSTM that is composed of the MPO-format and nonlinear structure while the previous methods use low-rank decomposition in part of an LSTM model. Second, the proposed MPO-LSTM model can be applied to an existing model directly, there is no definition of new layers.

To bridge the gap between the high-performance and the high-cost, many approaches have been proposed to compress such large, hyper-parametric neural networks, including parametric pruning and sharing \cite{gong2014compressing, huang2018learning}, low-rank matrix decomposition \cite{jaderberg2014speeding, zhao2021Semi}, and knowledge distillation \cite{hinton2015distilling, Fahad2022Dynamic}. But, most of these methods have been applied to feed-forward neural networks and convolutional neural networks (CNN), while little attention has been paid to compressing LSTM models \cite{belletti2018factorized, lu2016learning}, especially in NLP and speech enhancement tasks. 
Furthermore, these existing works rely on strong prior knowledge or large computational resources as well as impair model performance.
It is worth noting that \cite{see2016compression} applied the parameter pruning to the standard Seq2Seq \cite{sutskever2014sequence} architecture in neural machine translation, which adopts an  LSTM model for encoders and decoders. In addition, in language modeling, \cite{tjandra2017compressing} used tensor-train decomposition proposed by \cite{oseledets2011tensor} that is mathematically equivalent to the matrix product operators, \cite{wen2017learning} used the binarization technology, \cite{yang2017tensor} adopted the architectural changes to approximate the low-rank decomposition in part of LSTM model, and \cite{gao2020compressing} demonstrated that the matrix product operators(MPO) method is well effective for model compression by using the MPO to replace the linear transformation of the fully connected layer and the convolution layer. 
\cite{sun2020model} has also verified that it is very effective to compress a linear part of the network with the matrix product operators method on acoustic data sets.
In the study of lightweight fine-tuning of pre-trained models in natural language processing, \cite{liu2021enabling} proposes that the MPO approach can be used to distinguish the central tensor from the auxiliary tensor, and by fine-tuning a few auxiliary tensor parameters, an efficient and fast lightweight fine-tuning scheme can be achieved.

To the best of our knowledge, this is the first study that focused on compressing an LSTM model fully with the MPO-based representation.
We provide an in-depth analysis of why the MPO representation is ahead of the LSTM module in terms of expressiveness.
In principle, our method can also combine with other categories method, such as quantization \cite{courbariaux2015binaryconnect} and Huffman coding \cite{han2015learning}, for obtaining higher compression ratios. 
\section{Conclusion}\label{conclusion}
In this paper, we demonstrate that an LSTM model can be well compressed using the MPO method with proper orders and balanced dimensions of modes.
We also present the MPO-LSTM model based on our demonstration for LSTM model compression. 
The advantage of our methods over the pruning method is that we do not require recording the indices of nonzero elements.
In this method, we use the MPO decomposition format to replace the weight matrices in linear transformations in the LSTM models. 
We evaluate the models under different compression rates with several datasets. 
The experiment results on IMDB, SST and VBD show that our proposed MPO method well outperforms the pruning method in NLP problem and speech enhancement performance under the same compression rate for an LSTM model.
Thus, the MPO method can be applied as a simple and efficient pruning method.
In the future, the MPO-based model compression method can be used in many other tasks, and it is also an interesting problem to explore the combination of MPO method and other compression methods.
\section*{Acknowledgments}
This research is financially supported by the National Natural Science Foundation of China under Grants 11934020, 11722437, 11674352 and 11774422.

Ze-Feng Gao, Xingwei Sun and Lan Gao contributed equally to this work.

\bibliographystyle{apalike2}
\bibliography{ref}
\end{document}